\documentclass[10pt,journal,compsoc]{IEEEtran}
\usepackage[dvipsnames]{xcolor}
\usepackage[pdftex]{graphicx}
\usepackage[most]{tcolorbox}
\usepackage{hyperref}
\usepackage{listings}
\usepackage{booktabs}
\usepackage{subcaption}

\usepackage{xspace}
\usepackage{amsmath,amssymb}

\newcommand{\ie}{\textit{i.e.,}}
\newcommand{\eg}{\textit{e.g.,}}
\newcommand{\etal}{\textit{et al.}}

\newcommand{\github}{{\sc GitHub}}

\newcommand{\llm}{\textsc{LLM}}
\newcommand{\llms}{\textsc{LLM}s}
\newcommand{\codebert}{\textsc{CodeBERT}}
\newcommand{\graphcodebert}{\textsc{GraphCodeBERT}}
\newcommand{\codetfive}{\textsc{CodeT5}}
\newcommand{\unixcoder}{\textsc{UnixCoder}}
\newcommand{\csn}{\textsc{csn}}
\newcommand{\codesearchnet}{\textsc{CodeSearchNet}}
\newcommand{\se}{\textsc{SE}}
\newcommand{\ml}{\textsc{ML}}
\newcommand{\dl}{\textsc{DL}}
\newcommand{\codetrans}{\textsc{CodeTrans}}
\newcommand{\pydataset}{\textsc{Python-150}}
\newcommand{\tlcodesum}{\textsc{TL-CodeSum}}
\newcommand{\tlc}{\textsc{TLC}}
\newcommand{\funcom}{\textsc{Funcom}}
\newcommand{\fcm}{\textsc{FCM}}
\newcommand{\bigclonebench}{\textsc{BigCloneBench}}
\newcommand{\sourcerercc}{\textsc{SourcererCC}}
\newcommand{\deckard}{\textsc{Deckard}}
\newcommand{\codexglue}{\textsc{CodeXGlue}}
\newcommand{\bleu}{\textsc{BLEU}}
\newcommand{\mrr}{\textsc{MRR}}

\newcommand{\pvalue}{$p$-value}

\ifCLASSOPTIONcompsoc
  \usepackage[nocompress]{cite}
\else
  \usepackage{cite}
\fi
\ifCLASSINFOpdf
\else
\fi
\lstdefinestyle{mystyle}{
    language=Python,
    basicstyle=\ttfamily\small,
    commentstyle=\itshape\color{gray},
    keywordstyle=\bfseries\color{blue},
    frame=none,
    breaklines=true,
    breakatwhitespace=true,
    showstringspaces=false,
    tabsize=4,
}

\lstdefinestyle{mystylejava}{
    language=Java,
    basicstyle=\ttfamily\tiny,
    commentstyle=\itshape\color{gray},
    keywordstyle=\bfseries\color{blue},
    frame=none,
    breaklines=true,
    breakatwhitespace=true,
    showstringspaces=false,
    tabsize=4,
}

\begin{document}
%
\title{On Inter-dataset Code Duplication and \\Data Leakage in Large Language Models}
%
%
%
%

\author{José Antonio Hernández~López, Boqi Chen, Mootez Saad, Tushar Sharma, Dániel Varró
\IEEEcompsocitemizethanks{\IEEEcompsocthanksitem J.A. Hernández López and Dániel Varró are with IDA, Linköping University, Sweden.\protect\\
E-mail: \{jose.antonio.hernandez.lopez, daniel.varro\}@liu.se
\IEEEcompsocthanksitem Mootez Saad and Tushar Sharma are with Dalhousie University, Canada.\protect\\
E-mail: \{mootez, tushar\}@dal.ca
\IEEEcompsocthanksitem Boqi Chen is with McGill University, Canada.\protect\\
E-mail: boqi.chen@mail.mcgill.ca
}}

%
%

\markboth{Journal of \LaTeX\ Class Files,~Vol.~14, No.~8, August~2015}%
{Shell \MakeLowercase{\textit{et al.}}: Bare Advanced Demo of IEEEtran.cls for IEEE Computer Society Journals}
%



\IEEEtitleabstractindextext{%
\begin{abstract}
\textit{Motivation.} Large language models (\llms{}) have exhibited remarkable proficiency in diverse software engineering (\se{}) tasks, such as code summarization, code translation, and code search. 
Handling such tasks typically involves acquiring foundational coding knowledge on large, general-purpose datasets during a pre-training phase, and subsequently refining on smaller, task-specific datasets as part of a fine-tuning phase. 

\textit{Problem statement.} Data leakage \ie{} using information of the test set to perform the model training,
is a well-known issue in training of machine learning models. A manifestation of this issue is the intersection of the training and testing splits. While \textit{intra-dataset} code duplication examines this intersection within a given dataset and has been addressed in prior research, \textit{inter-dataset code duplication}, which gauges the overlap between different datasets, remains largely unexplored. If this phenomenon exists, it could compromise the integrity of \llm{} evaluations because of the inclusion of fine-tuning test samples that were already encountered during pre-training, resulting in inflated performance metrics.

\textit{Contribution.} This paper explores the phenomenon of inter-dataset code duplication and its impact on evaluating \llms{} across diverse \se{} tasks. 

\textit{Study design.} We conduct an empirical study using the \codesearchnet{} dataset (\csn{}), a widely adopted pre-training dataset, and five fine-tuning datasets used for various \se{} tasks. 
We first identify the intersection between the pre-training and fine-tuning datasets using a deduplication process. {\color{black}
Next, we pre-train two versions of \llms{} using a subset of \csn{}: one leaky \llm{}, which includes the identified intersection in its pre-training set, and one non-leaky \llm{} that excludes these samples. Finally, we fine-tune both models and compare their performances using fine-tuning test samples that are part of the intersection.} 

\textit{Results.} Our findings reveal a potential threat to the evaluation of \llms{} across multiple \se{} tasks, stemming from the inter-dataset code duplication phenomenon. We also demonstrate that this threat is accentuated by the chosen fine-tuning technique. {\color{black} Furthermore, we provide evidence that open-source models such as \codebert{}, \graphcodebert{}, and \unixcoder{} could be affected by inter-dataset duplication.} Based on our findings, we delve into prior research that may be susceptible to this threat. Additionally, we offer guidance to \se{} researchers on strategies to prevent inter-dataset code duplication.


\end{abstract}

}

\maketitle

\IEEEdisplaynontitleabstractindextext

%
\IEEEpeerreviewmaketitle

\section{Introduction}

Recent advances in machine learning (\ml{}) have significantly transformed the landscape of software development. \ml{} has increased the efficiency of software development by automating various software engineering (\se{}) tasks~\cite{Sharma2024}. For instance, \ml{} models have been trained to generate documentation from source code~\cite{allamanis2016convolutional,iyer2016summarizing,wei2019code}, a task often deemed tedious by software engineers~\cite{iyer2016summarizing}. These models have also improved code search~\cite{gu2018deep,husain2019codesearchnet,wang2020trans}, empowering developers to leverage existing source code for code reuse or comprehension of functionality~\cite{di2023code}. Furthermore, \ml{} models can be used for the automatic migration of source code from one programming language to another~\cite{lu2021codexglue}. 

In this context, many recent \ml{} solutions rely on large language models (\llms{})~\cite{feng2020codebert,guo2020graphcodebert,guo2022unixcoder,wang2021codet5}, which have exhibited remarkable performance in such tasks. \llm{}-based solutions are often realized through a prevalent pre-training/fine-tuning approach~\cite{feng2020codebert, guo2020graphcodebert,guo2022unixcoder,wang2021codet5}. Firstly, the model is \emph{pre-trained} using a large corpus of unlabeled source code. Subsequently, the model is \emph{fine-tuned} for a specific task using a more focused, smaller dataset. For example, \codebert{}~\cite{feng2020codebert} uses the \codesearchnet{}~\cite{husain2019codesearchnet} dataset (with  $\sim$6.4M code snippets) for pre-training, and it can be fine-tuned for code search using the \pydataset{} dataset~\cite{barone2017parallel,wan2022naturalcc,zhou2021assessing} (with $\sim$92K code snippets).

However, a significant proportion of the source code used as datasets in software engineering research originates from a single source \ie{} \github{}~\cite{barone2017parallel,lu2021codexglue,husain2019codesearchnet,hu2018summarizing,leclair2019neural}. For instance, the datasets in \codexglue{}~\cite{lu2021codexglue}, a widely utilized benchmark for code understanding, predominantly consist of samples extracted from \github{}. Moreover, there is a substantial amount of duplication  within \github{} as noted by various researchers~\cite{lopes2017dejavu,gharehyazie2019cross,golubev2021nature}. 

These observations may indicate a potential overlap between the datasets used for pre-training and fine-tuning in \llm{}-based solutions when applied to diverse \se{} tasks. We denote this phenomenon as \emph{inter-dataset code duplication}. 
Therefore, a first objective of this paper is to investigate the following research question:

\begin{tcolorbox}
    To what extent does \emph{inter-dataset code duplication} or, even worse, data leakage exist between pre-training and fine-tuning datasets used in \llm{}-based solutions for \se{}?
\end{tcolorbox}



This phenomenon can be interpreted as a special form of data leakage, where samples encountered during pre-training are also used in evaluating the fine-tuned model. As a result, a fine-tuned \llm{}-based solution may perform better on samples that have already been seen during pre-training, which influences (inflates) the overall performance thus compromising the integrity of performance evaluations. As a consequence, when such solutions are used in production, the performance of an \llm{}-based solution might be lower than expected.
Hence, the second objective of this study is to address the following research question:
\begin{tcolorbox}
Does \emph{inter-dataset code duplication} pose a threat to the validity of \llms{} evaluations?
\end{tcolorbox}


To the best of our knowledge, neither of these questions has been systematically investigated in past research. In fact, a recurring practice observed when using \llms{} for various \se{} tasks involves pre-training and fine-tuning with the same dataset (or fine-tuning with a subset of the pre-training dataset)~\cite{wang2021codet5,feng2020codebert,guo2020graphcodebert,guo2022unixcoder}. Consequently, a positive response to the research questions could imply that some findings in those works might report overly optimistic results due to the inter-dataset code duplication phenomenon.


{\color{black} To demonstrate the existence of inter-dataset code duplication, we first calculate the overlap between \codesearchnet{}~\cite{husain2019codesearchnet}, a widely-used open-source pre-training dataset, and five popular fine-tuning datasets: \codetrans{}\cite{lu2021codexglue}, \pydataset{}\cite{barone2017parallel}, \tlcodesum{}\cite{hu2018summarizing}, \funcom{}\cite{leclair2019neural}, and \bigclonebench{}~\cite{svajlenko2014towards}. Next, to show that this phenomenon poses a threat to the validity of \llms{} evaluations, we pre-train two versions of a given \llm{}: one \textit{leaky} \llm{}, whose pre-training set includes samples from the fine-tuning datasets, and one \textit{non-leaky} \llm{}, whose pre-training set excludes such samples. We then fine-tune and evaluate the performance of these models on the leaky portion of the test set for each \se{} task. Furthermore, we investigate whether the type of fine-tuning technique used might amplify the threat to validity. Beyond full fine-tuning, which modifies all model parameters, we assess three lightweight fine-tuning methods: LoRA~\cite{hu2021lora}, Prefix-Tuning~\cite{li2021prefix}, and layer freezing. Our rationale for considering these methods is twofold: 1) they enable fine-tuning with limited resources, making them more practical in certain cases~\cite{li2021prefix}, and 2) they introduce minimal modifications to the model's parameters, leading us to hypothesize that their evaluation may be more susceptible to the inter-dataset code duplication phenomenon.}

\noindent\textbf{Results.} Our results clearly confirm the existence of the duplication phenomenon, leading to non-trivial overlapping in three fine-tuning datasets. {\color{black} Additionally, we show that the leaky \llm{} performs better than its non-leaky counterpart over the leaky test portions. This evidences the existence of a threat to the \llms{} evaluations. Furthermore, we show that this threat is accentuated by the employed fine-tuning technique where lightweight fine-tuning techniques show higher impact. When using layer freezing, we observe an interesting property that applies only to the leaky \llm{} encoder (and not to the non-leaky \llm{}):
as the frozen layers increases, the performance of the model drops significantly for the non-leaky test portion compared to the leaky tests. We find the same pattern in the open-source models \codebert{}, \graphcodebert{}, and \unixcoder{}.}

Altogether, this work contributes to software engineering research in the following ways:
\begin{enumerate}
    \item We systematically demonstrate the presence of the inter-dataset code duplication phenomenon through comprehensive analysis using \csn{} and five state-of-the-art fine-tuning datasets.
    \item We provide evidence of a potential threat to the evaluations of \llms{} due to inter-dataset code duplication.
    \item {\color{black} We examine the effects of employing lightweight fine-tuning techniques (LoRA, prefix tuning, and layer freezing) versus full fine-tuning, in the context of the threat posed by inter-dataset code duplication.}
    \item {\color{black} We provide evidence that \codebert{}, \graphcodebert{}, and \unixcoder{} are vulnerable to this threat.}
\end{enumerate}


\noindent\textbf{Implications.} Our findings underscore the importance of carefully considering the dataset composition in the pre-training and fine-tuning phases of \llm{}-based solutions for \se{}. By acknowledging the potential for data leakage due to inter-dataset code duplication, researchers can enhance the robustness of their evaluation methodologies. This can be achieved by systematically excluding samples from the fine-tuning test set that also exist in the pre-training set. Additionally, our exploration delves into potential implications for previous \se{} research and establishes connections with earlier \ml{} research pertaining to catastrophic forgetting.

\noindent\textbf{Organization.} This paper is structured as follows. Firstly, Sect.~\ref{sect:background} overviews the background and related work. Sect.~\ref{sect:duplication} motivates inter-dataset code duplication and demonstrates its existence. Sect.~\ref{sect:leakage} evidences the threat to validity in the evaluation of \llms{} arising from this phenomenon. Finally, Sect.~\ref{sect:discussion} delves into a discussion of potential implications and connections with prior literature, while Sect.~\ref{sect:conclusion} concludes. 

{\color{black}
\noindent\textbf{Replication package.} The codebase for this paper is available at \url{https://github.com/Antolin1/code-inter-dataset-duplication}. The intersection between \codesearchnet{} and the five fine-tuning datasets is stored in a SQLite database and can be accessed at \url{https://zenodo.org/records/10446176}. Additionally, the leaky and non-leaky pre-trained \llms{} used in our experiments are available at \url{https://huggingface.co/antolin}.}


\section{Background \& Related Work}
\label{sect:background}

In this section, we provide key insights into the application of \llms{} in \se{}. We address aspects related to code duplication and data leakage, encompassing the definition of code clones, previously proposed clone detection tools, and the notion of \textit{code duplication} used in this paper. Additionally, we discuss previous studies about code duplication on \github{} and the impact of code duplication in \ml{} for~code.

\subsection{Large Language Models for Software Engineering}

\subsubsection{Pre-training and fine-tuning paradigm}
The initial deep learning (\dl{}) models introduced in the field of software engineering are tailored for specific tasks. In other words, these models are trained on a given dataset to execute a \se{} task by minimizing a designated loss function~\cite{allamanis2016convolutional,iyer2016summarizing,wei2019code}. However, with the introduction of \codebert{}~\cite{feng2020codebert}, there has been a shift in this paradigm towards pre-trained \llms{}. These models undergo initial pre-training on an extensive corpus of source code to acquire fundamental knowledge about both code and natural language. Subsequently, the model is fine-tuned to perform a specific task using a smaller dataset. Using this methodology, \llms{} have demonstrated impressive performance across various \se{} tasks, including code summarization~\cite{wang2021codet5}, code search~\cite{feng2020codebert}, code translation~\cite{lu2021codexglue}, code clone detection~\cite{lu2021codexglue}, and code generation~\cite{wang2023codet5+}, among others.


\subsubsection{Architecture of \llms{}}
\llms{} typically adopt the transformer architecture~\cite{vaswani2017attention} and can be categorized into three groups~\cite{min2023recent}: decoder-only, encoder-only, and encoder-decoder language models.

\textbf{Decoder-only language models} are normally trained to predict the next token given the preceding ones. In the context of code, examples of such models include \textsc{CodeGPT}~\cite{lu2021codexglue}, \textsc{StarCoder}~\cite{li2023starcoder}, \textsc{CodeGen}~\cite{nijkamp2022codegen}, etc. These models are typically used for code generation tasks.

\textbf{Encoder-only language models} are usually trained on the masked language modeling task which aims to predict a masked word given the rest of the sequence as context. Examples in this category include \codebert{}~\cite{feng2020codebert}, \graphcodebert{}~\cite{guo2020graphcodebert}, and \unixcoder{}~\cite{guo2022unixcoder}. These models excel in classification tasks or tasks requiring feature extraction (such as code search or code clone detection).

\textbf{Encoder-decoder language models} 
generate text based on predefined conditions. Examples include \codetfive{}~\cite{wang2021codet5}, \codetfive{}+\cite{wang2023codet5+}, and \textsc{PLBART}\cite{ahmad2021unified}. They are intended for conditional generation tasks like code summarization or generating code from a natural language description.

This paper focuses only on encoder-only and encoder-decoder language models.

\subsubsection{Fine-tuning techniques}

Fine-tuning methods are essential for \llms{} to adapt to specific tasks and have been widely used in the context of \se{}. Many \llms{}, such as \codebert{} and \codetfive{}, have demonstrated effectiveness for tasks such as code search, code summarization, and code translation \cite{feng2020codebert,wang2021codet5} using the full fine-tuning approach. Recently, there has been a growing interest in applying lightweight fine-tuning for \llms{} in \se{}~\cite{weyssow2023exploring, ayupov2022parameter, shi2023towards} given the growing size of such models. These methods achieve performance comparable to the full fine-tuning method on various tasks and even outperform it when the dataset is small~\cite{wang2022no}. 

This paper focuses on four popular fine-tuning techniques: 
\emph{full fine-tuning}~\cite{devlin2018bert}, \emph{layer freezing}~\cite{lee2019would}, \emph{low-rank adaptation (LoRA)}~\cite{hu2021lora}, and \emph{prefix tuning}~\cite{li2021prefix}.
Such methods have been used in \se{} research extensively~\cite{feng2020codebert,ayupov2022parameter,shi2023towards,weyssow2023exploring}.

\textbf{Full fine-tuning} initializes the model with parameters learned from pre-training and further trains it on the task-specific dataset~\cite{devlin2018bert}. This method is straightforward to implement and tends to perform well with large and diverse datasets. However, this technique becomes computationally intensive as the size of the \llm{} increases.

\textbf{Layer freezing}
fine-tuning method involves freezing several of the initial \llm{} layers
that are typically responsible for extracting features, while updating the remaining layers during training.
Such mechanism results in an efficient way to fine-tune a model. In many cases, fine-tuning a small subset of the layers can yield results comparable to those of full fine-tuning \cite{lee2019would}.

\textbf{LoRA}~\cite{hu2021lora} enhances fine-tuning efficiency by introducing low-rank weight matrices to pre-trained \llm{}, rather than updating all weights directly. During fine-tuning, only these low-rank matrices are updated. This approach also reduces the storage requirements for the fine-tuned model, as only the low-rank matrices representing the changes to the original weight matrix need to be stored.

\textbf{Prefix tuning}~\cite{li2021prefix} is a recent fine-tuning approach inspired by prompting methods for \llms{}. In this approach, continuous task-specific vectors, also known as prefixes, are prepended to each transformer block of the \llm{}. These prefixes are learned during fine-tuning while weights of the \llm{} remain frozen. The purpose of prefix tuning is to guide the \llm{}'s processing towards a task-specific dataset using these prefixes. This method is especially beneficial for small datasets, as it effectively prevents overfitting. 

In this paper, we conduct experiments with these four fine-tuning techniques to discern which of them is more impacted by the inter-dataset code duplication phenomenon.

\subsection{Code duplication}

\subsubsection{Code clones}
In the code clone literature, various types of code clones are identified~\cite{roy2007survey,sajnani2016sourcerercc,golubev2021nature}. Type-1 clones involve identical code snippets with differences in spaces, comments, whitespaces, or layouts. Type-2 clones are syntactically equivalent fragments with variations in identifiers, literals, and types, along with Type-1 differences. Type-3 clones encompass all changes found in Type-1 and Type-2 clones, with additional modifications like altered, added, or removed statements. Finally, Type-4 clones represent snippets that share functionality but have different syntax.


Code clone detection tools rely on the concept of \textit{similarity} to determine whether two code fragments are clones or not~\cite{zakeri2023systematic}. Each tool utilizes its notion of similarity. For instance, \sourcerercc{}~\cite{sajnani2016sourcerercc} measures this similarity by assessing the overlap between the tokens in two code fragments. In contrast, \deckard{}~\cite{jiang2007deckard} takes into account \textsc{AST}s and employs a tree edit distance metric.


\subsubsection{Notion of code duplication and tool employed}
In this paper, we adopt the concept of \emph{code duplication} as defined by Allamanis~\cite{allamanis2019adverse}, referring to \textit{code snippets that are highly similar but not necessarily identical}. That is, we are interested in Type-3 code clones. 
Particularly, we use a tool based on \sourcerercc{}~\cite{sajnani2016sourcerercc}
used by Allamanis~\cite{allamanis2019adverse}
to identify inter-dataset code duplication.
We adopt this concept of code duplication and the associated tool for two main reasons: 1) the author has conducted a similar study but at an intra-dataset level, 
and 2) the tool is highly efficient and fast.
The tool reports duplication between two code fragments if the Jaccard similarities of their two ``fingerprints'' cross a pre-defined threshold.
The first fingerprint comprises a multi-set of literals and identifiers, while the other consists of a set of literals and identifiers.


\subsection{Code duplication within GitHub}

\sourcerercc{}~\cite{sajnani2016sourcerercc} is used by Lopes \etal{}~\cite{lopes2017dejavu} to perform an extensive study of code duplication within \github{}~\cite{lopes2017dejavu}. The study analyzes 4.5 million \github{} projects, encompassing 428M files written in Python, Java, and JavaScript. Key findings reveal that 70\% of code on \github{} consists of clones of previously created files, and a noteworthy amount of inter-project duplication exists (\ie{} files in one project can be found in another project).

Gharehyazie \etal{}\cite{gharehyazie2019cross} employs \deckard{}\cite{jiang2007deckard} to identify code clones within 8k \github{} projects. Employing a statistical and network approach, they conclude that cross-project cloning is a common practice across various granularities, ranging from several lines to entire projects. Additionally, they observe an onion model of cloning within \github{}, where the majority of clones originate from the same project, followed by projects within the same domain, and then from projects in different domains.

In an empirical study on the nature of code clones, Golubev and Bryksin~\cite{golubev2021nature} investigate a corpus of 23k open-source Java projects. Their findings indicate that, in Java, method-level duplication is more prevalent than file-level duplication. They also highlight the common practice of copying/pasting code over the years of Java code existence.

Lastly, Yang \etal{}~\cite{yang2017stack} provide compelling evidence of code migration from \textsc{Stack Overflow} to \github{}, a phenomenon contributing to code duplication within \github{}. Using \sourcerercc{}, their study identifies correspondences between 290M function definitions within \github{} and 1.9M Python snippets from \textsc{Stack Overflow}.

The extensive body of research presenting a noteworthy occurrence of duplication within \github{}, whether inter- or intra-repository, along with the choice of \github{} as the primary source for code snippets in \ml{} datasets, emphasizes the significance of evaluating \emph{inter-dataset code duplication}.

\subsection{Code duplication and ML for code}
\label{sect:codedupml4code}

Motivated by this amount of duplication within \github{}, Allamanis~\cite{allamanis2019adverse} conducts a study on \emph{intra-dataset code duplication}, specifically examining the intersection between training and testing splits within code datasets and its implications for \ml{} models. The study demonstrates that the performances of \ml{} models are inflated as a result of the non-trivial intersection. Additionally, the author presents a set of best practices for collecting code corpora and evaluating machine learning models on them. Subsequently, researchers, building upon this knowledge, have been aware of duplication issues and commonly implement deduplication strategies in their datasets~\cite{karmakar2023jemma,husain2019codesearchnet,izadi2022codefill,kocetkov2022stack}. 
For example, Husain \etal{}\cite{husain2019codesearchnet} implement a deduplication procedure before releasing the \codesearchnet{} dataset. Karmakar \etal{}\cite{karmakar2023jemma} incorporate deduplication as part of the preprocessing pipeline before releasing \textsc{JEMMA}, a large-scale, high-quality Java code corpus. Kocetkov \etal{}~\cite{kocetkov2022stack} release the deduplicated version of the \textsc{Stack}, a massive 6TB source code corpus, demonstrating that pre-training code \llms{} with a deduplicated version yields better results.


Our study differs from the Allamanis's work~\cite{allamanis2019adverse}. While this prior research focuses on \emph{intra-dataset code duplication}, our paper explores the \emph{inter-dataset code duplication}, which poses a new potential threat to the evaluation of \ml{} models given the recent shift in learning paradigms (task-specific training to pre-training/fine-tuning).

\subsection{Data leakage in \llms{} for code}

Sallou \etal{}~\cite{sallou2023breaking} addresses the issue of data leakage in \llms{} for code.
Through interactions with \textsc{ChatGPT}, the authors show that \textsc{ChatGPT} is aware of multiple samples from the \textsc{Defects4J} dataset~\cite{just2014defects4j}. {\color{black} Furthermore, Lee \etal{}~\cite{lee2023github} demonstrate that there exists a significant overlap between the \textsc{Defects4J} dataset and the \textsc{Stack}~\cite{kocetkov2022stack}. This implies that \llms{} such as \textsc{StarCoder} have already encountered samples from that dataset during pre-training. Consequently, \textsc{Defects4J} may not be a reliable dataset for evaluating current \llms{}. To address this issue, Lee \etal{} propose a new dataset comprising 76 real-world Java bugs collected after the data snapshot of ChatGPT and GPT-4.}

Additionally, Schafer \etal{}~\cite{schafer2023empirical} observe that \textsc{StarCoder}\cite{li2023starcoder}, along with other similar models like \textsc{Codex}~\cite{chen2021evaluating}, exhibits improved performance in test generation when using \textsc{HumanEval} samples (available at \github{}) compared to \textsc{SF110} samples (available at Sourceforge but not at \github{}~\cite{sallou2023breaking}). Notably, \textsc{StarCoder} was trained on the \textsc{Stack}~\cite{kocetkov2022stack}, a dataset extracted from \github{}.

This paper's contribution differs from previous works in that it explores data leakage within a pre-training/fine-tuning framework and analyzes factors that may exacerbate the threat to the validity of the LLMs evaluations. Previous literature has primarily demonstrated data leakage by eliciting responses from \llms{} through various prompts.

\section{Inter-dataset code duplication}
\label{sect:duplication}

In this section, we motivate the need to assess the inter-dataset code duplication phenomenon. Subsequently, we detail the study design and showcase the results that demonstrate the existence of inter-dataset code duplication.

\subsection{Motivation}

The inter-dataset code duplication phenomenon refers to the overlapping in terms of duplicated code samples among code datasets. In conventional \ml{} scenarios, this phenomenon is not problematic since each dataset is used independently for training and evaluating \ml{} models. However, issues arise when employing a pre-training/fine-tuning approach, particularly when evaluating test samples encountered during pre-training. This scenario has the potential to inflate performance metrics, thereby compromising the \llm{} evaluation and the estimation of its performance in production. To illustrate this concern further, let us examine the task of generating natural language descriptions based on a given code snippet (\ie{} the code summarization task). For this task, we opted to fine-tune \codetfive{}~\cite{wang2021codet5} using the \pydataset{} dataset~\cite{barone2017parallel}. Consider the Python test sample shown in Listing~\ref{lst:intro}.

\lstset{style=mystyle}
\begin{lstlisting}[caption={Example extracted from the \pydataset{} test set.}, label={lst:intro}]
# round a to the nearest integer 
# if that integer is within an epsilon of a.
def round_if_near_integer(a, epsilon=0.0001):
    if (abs((a - round(a))) <= epsilon):
        return round(a)
    else:
        return a
\end{lstlisting}

\noindent This test sample is also part of the \codesearchnet{} dataset~\cite{husain2019codesearchnet}, used for pre-training \codetfive{}. Therefore, \codetfive{} is already familiar with this sample. Additionally, a pre-training task closely resembling the fine-tuning task (\textit{the bimodal dual generation task}~\cite{wang2021codet5}) is performed in the pre-training phase of \codetfive{}. This example is equally applicable to other models that undergo pre-training using the \codesearchnet{} dataset. For example, the same concern emerges when examining the \unixcoder{} model~\cite{guo2022unixcoder} in the context of semantic code search. Consequently, the performance of these models on such samples could be inflated, leading to incorrect conclusions in the evaluation of \llms{} and inferior performance of such \llms{} in production.

\subsection{Study design}

This section outlines the experimental setup adopted to demonstrate the presence of this phenomenon. Specifically, we introduce the datasets used and an explanation of the approach for computing the inter-dataset duplicates.

\subsubsection{Datasets}

\begin{figure*}[h]
    \centerline{\includegraphics[width=0.7\textwidth]{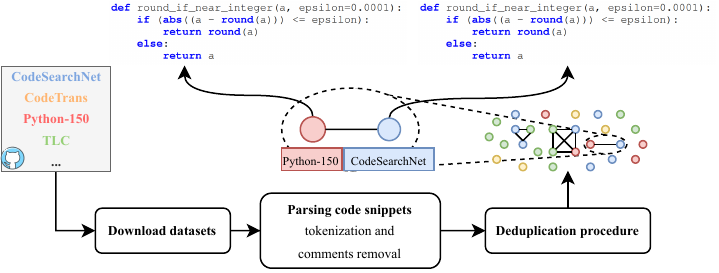}}
    \caption{Our approach to assessing inter-dataset code duplication.}
    \label{fig:approach}
\end{figure*} 

The datasets analyzed in this paper are the following:
\\\\
\noindent\textbf{The \codesearchnet{} dataset (\csn{})}~\cite{husain2019codesearchnet} provides a comprehensive multilingual dataset with 6.4M code snippets, approximately 2M of which are accompanied by natural language documentation. The dataset encompasses snippets from six programming languages, with our focus limited to Java and Python. Several language models, including \codetfive{}, \codebert{}, \graphcodebert{}, and \unixcoder{}, are pretrained using this dataset.
\\\\
\noindent\textbf{The \codetrans{} dataset}~\cite{lu2021codexglue} serves as a fine-tuning dataset composed by $\sim$12k pairs of Java and C\# methods. This dataset is tailored to code translation which aims to migrate legacy software from one programming language in a platform to another programming language.
\\\\
\noindent\textbf{The \pydataset{} dataset}\cite{barone2017parallel} consists of pairs comprising Python functions and their corresponding documentation. For our study, we use the version accessible through \textsc{NaturalCC}~\cite{wan2022naturalcc}, encompassing  $\sim$92k pairs. Originally intended for code summarization, this dataset has also been employed in prior research for the code search task~\cite{zhou2021assessing}.
\\\\
\noindent\textbf{The \tlcodesum{} (\tlc{}) and \funcom{} (\fcm{}) datasets}~\cite{hu2018summarizing,leclair2019neural} consist of 87k and 2M of Java methods and their corresponding documentation, respectively. These datasets are specifically curated for code summarization.
\\\\
\noindent\textbf{The \bigclonebench{} dataset}~\cite{svajlenko2014towards} is a comprehensive dataset designed for code clone detection. It contains a substantial collection of Java code clones, providing a valuable resource for evaluating the similarity between code fragments. We use the filtered version available in \codexglue{}\cite{lu2021codexglue}, including around 9k code snippets.

\subsubsection{Approach}

Our method for revealing the existence of inter-dataset duplication is depicted in Fig.~\ref{fig:approach}, including three key steps:
\\\\
\noindent \textbf{Dataset Acquisition.} The datasets are obtained following the train/test/valid splits established in previous research. For the \codetrans{} and \bigclonebench{} datasets, we adhere to the splits provided by the \codexglue{} benchmark~\cite{lu2021codexglue}. Regarding the \pydataset{} dataset, we use the splits presented in \textsc{NaturalCC}~\cite{wan2022naturalcc}. For \tlc{}, the original splits are employed~\cite{hu2018summarizing}. Lastly, for \fcm{}, we adopt the splits utilized by Shi \etal{}~\cite{shi2022evaluation}.
\\\\
\noindent \textbf{Parsing the code snippets.} Each code snippet undergoes parsing to extract the code tokens, excluding comments. We use the Python \texttt{tokenize} module for Python code as it is included in the Python standard library\footnote{\url{https://docs.python.org/3/library/tokenize.html}}. For Java, we use the \texttt{javalang} library\footnote{\url{https://github.com/c2nes/javalang}}, a Python library for parsing Java source code used in previous \se{} works~\cite{zhang2022astro,li2017software}.
\\\\
\noindent \textbf{Running the deduplication procedure.} After parsing all snippets from the datasets, a deduplication procedure is executed. We employ the tool developed by Allamanis~\cite{allamanis2019adverse}, which is based on \textsc{SourcererCC}~\cite{sajnani2016sourcerercc}. 
The thresholds applied in this study to identify a pair of code snippets as clones adhere to the recommendations of the tool's creator~\cite{allamanis2019adverse} ($0.7$ for the multiset and $0.8$ for the set).
\\\\
The ultimate result of our pipeline is represented by an undirected sparse graph. In this graph, nodes correspond to individual code snippets, and edges indicate instances where one code snippet duplicates another. For instance, the node corresponding to the snippet in Listing~\ref{lst:intro} is linked to another node representing its duplicate in the \csn{} dataset. To facilitate subsequent analysis and research, we provide this graph in the form of an SQLite database\footnote{Available at \url{https://zenodo.org/records/10446176}}.

\subsection{Results}

Using \csn{} as the pre-training dataset, we compute the inter-dataset duplication percentage ($IDD_{\mathcal{D}}$) for each fine-tuning dataset $\mathcal{D}$.
\begin{equation}
    IDD_\mathcal{D} = \frac{|\mathcal{D} \cap \text{\csn{}}|}{|\mathcal{D}|}\times100\%
\end{equation}
In this expression, the numerator represents the count of samples belonging to $\mathcal{D}$ that contain at least one duplicate in \csn{} (\textit{i.e.}, one of its neighbors within the undirected graph is part of \csn{}).

\begin{figure}[tb]
    \centerline{\includegraphics[width=0.4\textwidth]{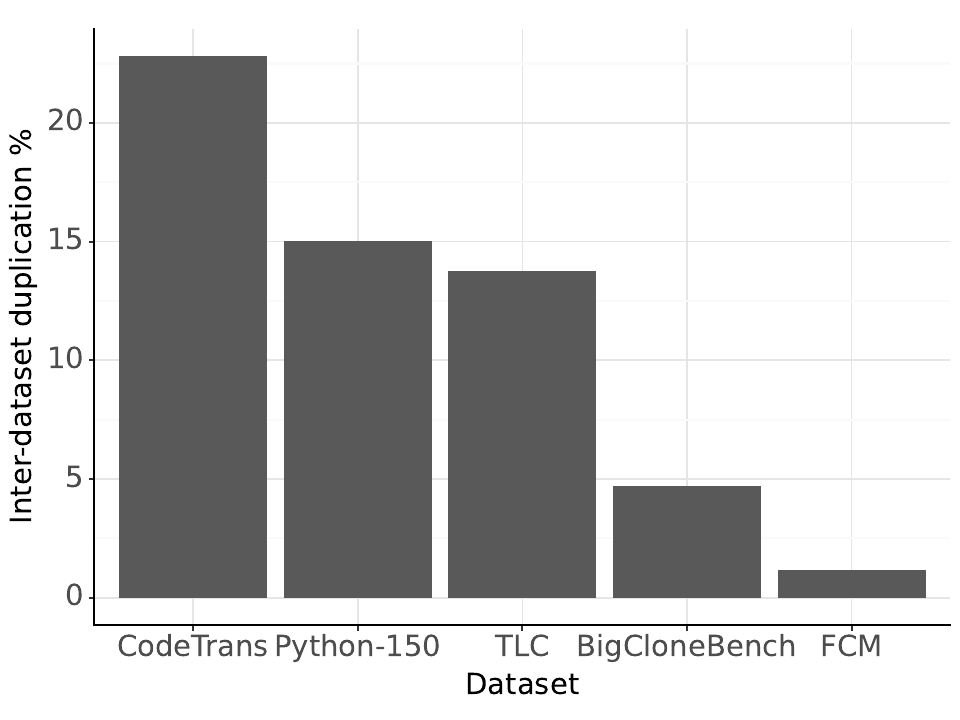}}
    \caption{Inter-dataset duplication wrt. \csn{} for each dataset.}
    \label{fig:interdup}
\end{figure}

The barplot in Fig.~\ref{fig:interdup} illustrates the inter-dataset duplication percentages for each dataset. Particularly notable are the \codetrans{}, \pydataset{}, and \tlc{} datasets, which display a substantial level of inter-dataset duplication (with ratios ranging between 13\% and 23\%) with \csn{}. In contrast, the inter-dataset duplication percentages for the \fcm{} and \bigclonebench{}\footnote{The $IDD_\mathcal{D}$ for this dataset was calculated based on the collection of unique code samples.} datasets are notably below 5\%.

\begin{tcolorbox}
   Our results reveal notable duplication percentages, ranging from 13\% to 23\%, in three datasets (\tlc{}, \pydataset{}, and \codetrans{}) with respect to \csn{}. This observation confirms the presence of the inter-dataset code duplication phenomenon in datasets employed for \llm{}-based solutions.
\end{tcolorbox}

\section{The effect of the inter-dataset duplication on \llms{} evaluation}
\label{sect:leakage}

The existence of the inter-dataset duplication phenomenon,
specifically data leakage,
implies the incorporation of samples from the pre-training phase into the test set of fine-tuning datasets. This could compromise the evaluation of the \llms{} as the fine-tuned model may perform better on those samples. As a result, this section investigates the potential impact of such data leakage on the evaluation of \llms{} in various \se{} tasks.

\subsection{Study design}
{\color{black}In this section, we first introduce the code tasks used in the study, followed by the construction of the pre-training datasets and the considered \llms{} to be pre-trained. Next, we describe the approach used to demonstrate the presence of the threat to \llms{} evaluation. Finally, we present the experimental settings.}

\subsubsection{Datasets, code tasks, and evaluation metrics}

We employ three datasets with a substantial inter-dataset duplication percentage to reliably measure the impact: \codetrans{}, \pydataset{}, and \tlc{}. Table~\ref{tab:percen} shows the duplication percentage with respect to \csn{} for each dataset as well as the duplication percentage within the test set concerning \csn{}, and the associated tasks. Consequently, the targeted code tasks in this paper are listed below.

\begin{itemize}
    \item \emph{Code translation}. Employing \codetrans{} as the target dataset, we address the Java-to-C\# code translation task. The evaluation metric used is the smoothed \bleu{} score~\cite{papineni2002bleu}, as utilized in prior studies~\cite{lu2021codexglue}. This metric takes values in the range of $[0, 100]$ (the higher the better) and measures the similarity of the generated code to the ground truth code through $n$-gram (contiguous sequences of $n$ tokens) overlapping.
    \item \emph{Code summarization}. Utilizing the \pydataset{} and \tlc{} datasets as the target datasets, we tackle the task of generating natural language descriptions based on given code snippets. The chosen evaluation metric is again the smoothed \bleu{} score~\cite{papineni2002bleu} between the generated summary and the ground truth; 
    it is a commonly employed metric for evaluating the code summarization task~\cite{feng2020codebert,lu2021codexglue}.
    \item \emph{Code search}. Using the \pydataset{} dataset, we address the challenge of retrieving code snippets that match the intent of a natural language query. For this task, our evaluation metric is the Mean Reciprocal Rank (\mrr{}), a metric used in previous work to evaluate the performance of code retrieval systems~\cite{lu2021codexglue,feng2020codebert,husain2019codesearchnet}. This metric is computed as the average of the reciprocal ranks (inverse of the position of the retrieved code snippet in the ranked list) and takes values in the range of $[0, 100]$ (the higher the better). We compute this metric using the ground truth snippet and 999 distractor code snippets~\cite{husain2019codesearchnet}.
\end{itemize}

\begin{table}[ht]
\caption{Percentages of inter-dataset duplication in comparison to \csn{} for both the entire dataset and the test split and the \se{} tasks enabled.}
\label{tab:percen}
\centering
\scalebox{0.85}{
\begin{tabular}{lccc}
\toprule
Dataset    & \multicolumn{1}{l}{\% full dataset} & \multicolumn{1}{l}{\% test set} & \se{} tasks                      \\ \hline
\codetrans{}  & 22.82\%                             & 22.40\%                         & Code translation             \\
\pydataset{} & 15.01\%                             & 14.79\%                         & Code search and summarization \\
\tlc{}        & 13.76\%                             & 13.56\%                         & Code summarization            \\ \bottomrule
\end{tabular}
}
\end{table}

\subsubsection{Pre-training datasets}

\begin{figure}[tb]
    \centerline{\includegraphics[width=0.47\textwidth]{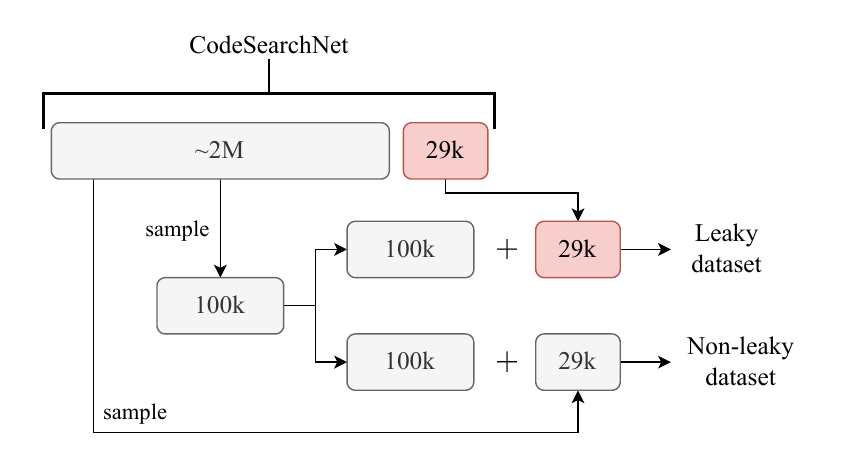}}
    \caption{Pre-training dataset construction. The red split within \codesearchnet{} represents the portion that is included in the fine-tuning datasets.}
    \label{fig:pretrainingcons}
\end{figure}

{\color{black} We construct two pre-training datasets: one that overlaps with the fine-tuning datasets (\emph{leaky dataset}) and another that remains completely disjoint (\emph{non-leaky dataset}).
Fig.~\ref{fig:pretrainingcons} outlines how these pre-training datasets were constructed. 

For the \emph{leaky dataset}, we first sample 100k code snippets from \csn{}. We then include the 29k snippets from \csn{} that overlap with all the fine-tuning datasets (indicated by the red split in the figure). This results in a pre-training dataset of approximately 129k snippets with data leakage. For the \emph{non-leaky dataset}, we start with the same initial 100k code snippets and sample an additional 29k snippets from \csn{} that neither appear in the fine-tuning datasets and nor intersect with the initial 100k samples. During  sampling, we ensure that the proportion of Java and Python code snippets remains consistent across both pre-training datasets. We use 100k snippets as the sample size to ensure the pre-training procedures remain manageable within our resource constraints.

As a result, we obtain two pre-training datasets of same size (each with approximately 129k snippets) and same proportion of Java and Python snippets.} 

\subsubsection{Pre-trained large language models}

{\color{black} Given the nature of the fine-tuning tasks, we pre-train two families of language models from scratch.

\noindent\textbf{For the code translation and code summarization tasks,} we pre-train encoder-decoder models. We use the \codetfive{} base architecture~\cite{wang2021codet5} along with its pre-trained vocabulary. Furthermore, we pre-train such models by employing the bimodal dual generation task used in \codetfive{} pre-training.

\noindent\textbf{For the code search task,} we pre-train encoder-only models. We use the \unixcoder{} architecture~\cite{guo2022unixcoder} and its pre-trained vocabulary. 
We employ the masked language modeling task as the pre-training task, which has been used in the pre-training of \codebert{}, \graphcodebert{}, and \unixcoder{}~\cite{feng2020codebert,guo2020graphcodebert,guo2022unixcoder}.
}

\subsubsection{Assessing the effect of the inter-dataset duplication}

{\color{black} Given one fine-tuning dataset, we have access to two \llms{}: one pre-trained with the leaky dataset (referred to as \textit{leaky LLM}) and the other with the non-leaky dataset (referred to as \textit{non-leaky LLM}). Both \llms{} will be fine-tuned using the same fine-tuning dataset. These fine-tuned \llms{} are then evaluated on the portion of the test set included in the leaky pre-training dataset (referred to as \textit{leaky test portion}).

If the \llm{} pre-trained on the leaky dataset outperforms the one pre-trained on the non-leaky dataset on this leaky test portion, it indicates that the inter-dataset code duplication poses a threat to the validity of \llm{} evaluations. This is because the better performance of the leaky \llm{} can be attributed to its prior exposure to these samples during pre-training. Moreover, the performance difference between the two \llms{} on this portion of the test set quantifies the impact of the inter-dataset duplication threat.} 
\subsubsection{Experimental settings}
\label{experimental_setting}

{\color{black}\noindent\textbf{Pre-training.} All the models (\ie{} the leaky and non-leaky variants of the encoder-only and encoder-decoder models) are pre-trained for 20 epochs with a batch size of 16 and an initial learning rate of $5e-5$ using the \textit{Adam} optimizer.}

\noindent\textbf{Full fine-tuning.}
We apply a consistent set of hyper-parameters across all language models, including learning rate and batch size. 
While slight adjustments are made in some models to enhance stability during fine-tuning or accommodate GPU constraints, an interested reader may find all the hyper-parameters used in our experiments in our repository\footnote{\url{https://github.com/Antolin1/code-inter-dataset-duplication}}.

{\color{black}
\noindent\textbf{LoRA and prefix tuning.} 
Following preliminary experimentation, we establish fixed parameters for LoRA, setting $r=8$, $\alpha=16$, and a dropout rate of $0.1$ across all scenarios. The target modules in the encoder-decoder models are $q$, $v$, $o$, and $k$, while for the encoder models, the target modules were $q$ and $v$. For prefix tuning, we set $20$ virtual tokens with prefix projection layers across all scenarios.
}

{\color{black}\noindent\textbf{Different seeds and statistical tests.} In general, each model is fine-tuned ten times with ten different seeds. However, in the layer-wise analysis (Sect.~\ref{sec:layerwise1} and Sect.~\ref{sec:layerwise2}), only five seeds are used. We then compare the performance of the leaky \llm{} and its non-leaky counterpart on the leaky portion of the fine-tuning test set by using the Mann–Whitney U test. We apply a significance level of $\alpha=0.05$.}


\noindent\textbf{Intra-dataset duplication.}
Note that we have performed (\textit{intra}-dataset) deduplication on the fine-tuning datasets to eliminate the potential impact of any train/test intersection in the results. 

\subsection{Results}

\subsubsection{Does inter-dataset code duplication pose a threat to the validity of \llms{} evaluations?}
\label{sect:integrity}

Due to inter-dataset code duplication, certain samples in the fine-tuning test set have been part of the pre-training process for the \llm{}. This study delves into whether this circumstance poses a threat to the validity of the
evaluation of \llm{}-based solutions for \se{}.

\begin{table}[]
\caption{Average score ($\pm$ standard deviation) for each type of \llm{} and for each dataset and task. The symbol \textdagger{} indicates that the differences between the performances of the leaky \llm{} and its non-leaky version are significant.}
\label{tab:firstTable}
\centering
\begin{tabular}{lcc}
\hline
                         & Leaky \llm{}   & Non-leaky \llm{} \\ \hline
\codetrans{}                & 87.56$\pm$0.56   & 87.41$\pm$1.21     \\ \hline
\pydataset{} (code sum.)   & 19.23$\pm$0.31   & 19.06$\pm$0.25     \\
\tlc{}                      & 18.80$\pm$ 0.41\textsuperscript{\textdagger}  & 17.55$\pm$0.44     \\ \hline
\pydataset{} (code search) & 79.04$\pm$0.31\textsuperscript{\textdagger} & 78.31$\pm$0.46   \\ \hline
\end{tabular}
\end{table}

{\color{black} Table~\ref{tab:firstTable} shows the performance of leaky and non-leaky \llms{} on the leaky test portion. In all cases, the leaky \llm{} outperforms the non-leaky \llm{}. For example, the average BLEU score of the leaky encoder-decoder over \tlc{} is 18.80, compared to 17.55 for the non-leaky counterpart. In other words, the \llm{} that was exposed to the test portion during pre-training performs better, on average, over that test portion than the \llm{} that did not encounter these samples during pre-training in the case of \tlc{}. 
However, we only observe statistical significance across two tasks out of four which are \pydataset{}-code search and \tlc{}. We hypothesize that this is attributed to the fine-tuning technique used. In this first experiment, we fully fine-tuned the models on the SE task, which updates all model parameters and potentially \textit{perturbs} the memorization that occurred during the pre-training phase. This hypothesis will be further examined in subsequent sections.
}

\begin{tcolorbox}
    {\color{black}
    We find that the fine-tuned leaky \llms{} achieve higher average performance compared to their non-leaky counterparts on the leaky test portion in all SE tasks and datasets.}
\end{tcolorbox}

\subsubsection{What is the effect of using lightweight fine-tuning techniques?}
\label{sect:otherfine}
{\color{black} Full fine-tuning involves updating all model parameters, which requires substantial GPU resources. Alternative methods have been introduced to fine-tune \llms{} more efficiently. Notable examples include Low-Rank Adaption (LoRA)~\cite{hu2021lora}, which updates low-rank weight matrices instead of all weights, and prefix tuning~\cite{li2021prefix}, which keeps the model parameters frozen and updates continuous task-specific vectors added at the beginning of the input. Given that these lightweight fine-tuning techniques tend to better preserve the knowledge acquired during pre-training, we hypothesize that such methods, which introduce minimal modifications to model parameters, could potentially increase the threat of inter-dataset code duplication.

To examine this hypothesis, we perform experiments by fine-tuning both \llms{}, the leaky \llm{} and its non-leaky counterpart using LoRA and prefix tuning. We then compare the outcomes with the results of the full fine-tuning approach.

Table~\ref{tab:secondTable} presents the performance of leaky and non-leaky \llms{} on the leaky test portion for each fine-tuning technique and dataset. In all cases, the comparison between the two versions of the \llm{} shows statistical significance. This contrasts with the previous section (\ie{} full fine-tuning), where statistical significance was observed in two (among four) cases.

Table~\ref{tab:effectsizes} presents the mean difference ($\Delta\mu$) and Cohen's $d$ effect size (standardized mean difference) between the leaky and non-leaky \llms{} for each fine-tuning technique, including full fine-tuning. For \codetrans{} and both datasets of \pydataset{}, LoRA and Prefix tuning achieve higher $\Delta\mu$ and $d$ compared to full fine-tuning. In contrast, for \tlc{}, full fine-tuning results in a higher $\Delta\mu$. However, when normalizing by the pooled standard deviation, LoRA and Prefix tuning yield higher $d$. This is due to the higher variances across the seeds observed in the full fine-tuning method.

One interpretation of these results is that using a lightweight fine-tuning technique, rather than modifying all parameters with full fine-tuning, increases the likelihood that the leaky \llm{} will outperform its non-leaky counterpart in the leaky test portion.

}

\begin{table*}[]
\caption{Average score ($\pm$ standard deviation) for each type of \llm{}, fine-tuning technique, and for each dataset and task. The symbol \textdagger{} indicates that the differences between the performances of the leaky \llm{} and its non-leaky version are significant.}
\label{tab:secondTable}
\centering
\begin{tabular}{lcc|cc}
\hline
                         & LoRA leaky \llm{} & LoRA non-leaky \llm{} & Prefix leaky \llm{} & Prefix non-leaky \llm{} \\ \hline
\codetrans{}                & 83.21$\pm$0.90\textsuperscript{\textdagger}      & 82.25$\pm$0.87          & 70.02$\pm$1.03\textsuperscript{\textdagger}        & 58.77$\pm$1.08            \\ \hline
\pydataset{} (code sum.)   & 17.84$\pm$0.22\textsuperscript{\textdagger}       & 16.98$\pm$0.16          & 17.21$\pm$0.13\textsuperscript{\textdagger}         & 16.02$\pm$0.15            \\
\tlc{}                      & 14.61$\pm$0.22\textsuperscript{\textdagger}      & 13.86$\pm$0.17         & 14.10$\pm$0.17\textsuperscript{\textdagger}         & 13.15$\pm$0.14            \\ \hline
\pydataset{} (code search) & 77.84$\pm$0.32\textsuperscript{\textdagger}      & 76.62$\pm$0.38          & 75.07$\pm$0.51\textsuperscript{\textdagger}        & 73.86$\pm$0.26            \\ \hline
\end{tabular}
\end{table*}

\begin{table}[]
\caption{Mean differences ($\Delta\mu$) and effect sizes ($d$) are presented for each fine-tuning technique and dataset. The symbol {\color{OliveGreen}$\uparrow$} indicates that the lightweight fine-tuning method shows a higher $\Delta\mu$ or $d$, while {\color{red}$\downarrow$} indicates a lower value with respect to full fine-tuning.}
\label{tab:effectsizes}
\centering
\scalebox{0.8}{
\begin{tabular}{lcccccc}
\hline
                         & \multicolumn{2}{c}{Full fine-tuning} & \multicolumn{2}{c}{LoRA} & \multicolumn{2}{c}{Prefix} \\
                         & $\Delta\mu$        & $d$        & $\Delta\mu$  & $d$  & $\Delta\mu$   & $d$   \\ \hline
\codetrans{}                & 0.14               & 0.15            & {\color{OliveGreen}$\uparrow$} 0.96         & {\color{OliveGreen}$\uparrow$} 1.08      & {\color{OliveGreen}$\uparrow$} 11.25         & {\color{OliveGreen}$\uparrow$} 10.63      \\ \hline
\pydataset{} (code sum.)   & 0.17               & 0.62            & {\color{OliveGreen}$\uparrow$} 0.86         & {\color{OliveGreen}$\uparrow$} 4.42      & {\color{OliveGreen}$\uparrow$} 1.19          & {\color{OliveGreen}$\uparrow$} 8.33       \\
\tlc{}                      & 1.25               & 2.94            & {\color{red}$\downarrow$} 0.75         & {\color{OliveGreen}$\uparrow$} 3.85      & {\color{red}$\downarrow$} 0.95          & {\color{OliveGreen}$\uparrow$} 6.00       \\ \hline
\pydataset{} (code search) & 0.73               & 1.89            & {\color{OliveGreen}$\uparrow$} 1.22         & {\color{OliveGreen}$\uparrow$} 3.48      & {\color{OliveGreen}$\uparrow$} 1.21          & {\color{OliveGreen}$\uparrow$} 3.01       \\ \hline
\end{tabular}
}
\end{table}


\begin{tcolorbox}
{\color{black} The results indicate that leaky \llms{} trained with lightweight fine-tuning methods consistently outperform their non-leaky counterparts on the leaky test portion. Additionally, these efficient techniques are more susceptible to the inter-dataset code duplication threat compared to the full fine-tuning method.} 
\end{tcolorbox}

\subsubsection{What is the effect of freezing the model?}
\label{sec:layerwise1}

\begin{figure}[t]
    \centerline{\includegraphics[width=0.45\textwidth]{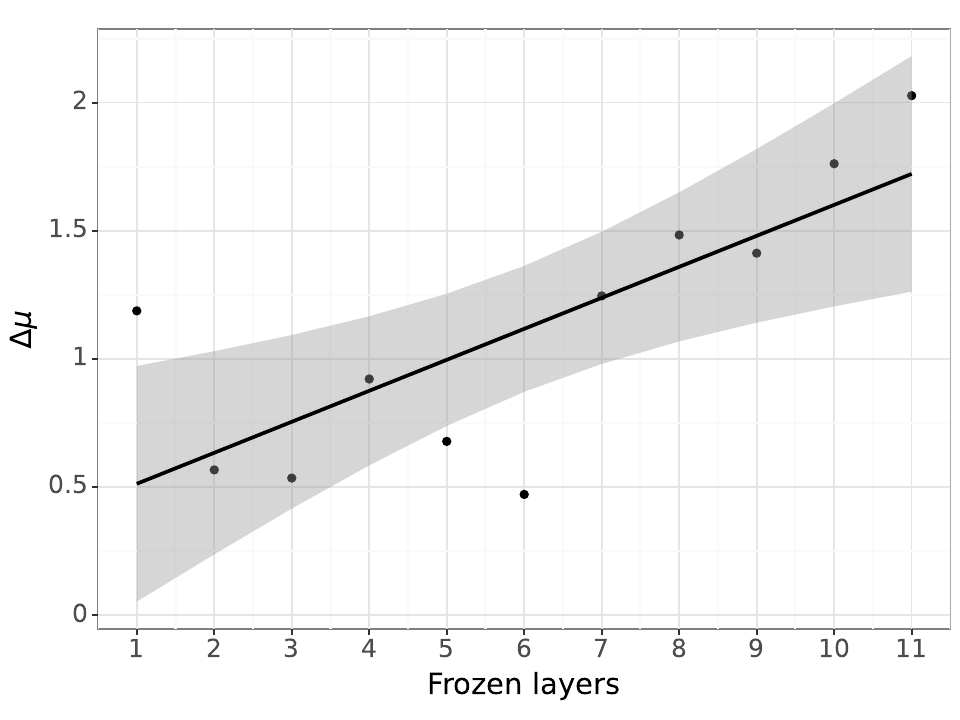}}
    \caption{Correlation plot where the $x-$axis is the number of frozen layers and the $y-$axis is the mean difference between the leaky and the non-leaky \llms{}.}
    \label{fig:corr}
\end{figure}

%
{\color{black}
Another way of fine-tuning the models without requiring substantial resources is to freeze model parameters~\cite{shi2023towards}. Similarly to LoRA and prefix tuning, our hypothesis is that this technique is prone to the threat of inter-dataset code duplication since it involves no alterations to an important subset of the model parameters, allowing the knowledge acquired during the pre-training phase to remain unchanged.

To test this hypothesis, we analyze the mean differences ($\Delta\mu$) between the leaky and non-leaky \llms{} when freezing their layers, concentrating exclusively on encoder-only models. We opt for this given the simplicity of layer-wise freezing compared to encoder-decoder models, where intricate interconnections exist between the encoder and decoder~\cite{vaswani2017attention}. In our experiments, we focus on the code search task. We freeze the lower $k$ transformer layers (of both the leaky and non-leaky \llms{}), specifically the $k$ layers nearest to the input, ranging from $1$ to $11$\footnote{Our pre-trained encoders follow the \unixcoder{} architecture, which consists of 12 transformer layers}, and measure the performance gaps between the leaky and non-leaky \llms{} over the leaky test portion.

Fig.~\ref{fig:corr} illustrates the correlation plot, revealing a positive correlation between the number of frozen layers and $\Delta\mu$. Both, the Spearman and Pearson coefficients show statistically significant correlation coefficients (\pvalue{}s~$<0.01$) of $0.75$ and $0.76$ respectively.
}

\begin{tcolorbox}
{\color{black}
A positive correlation is observed between the mean difference of the leaky \llm{} and its non-leaky counterpart, and the number of frozen layers. Consequently, freezing more parameters makes the \llm{} more susceptible to the effect of inter-dataset code duplication.
}
\end{tcolorbox}

\subsubsection{Assessing inter-dataset code duplication in open-source \llms{}}
\label{sec:layerwise2}

{\color{black}
In our previous experiments (Sections~\ref{sect:integrity}, \ref{sect:otherfine}, and \ref{sec:layerwise1}),
by controlling the pre-training procedure,
we demonstrate that inter-dataset code duplication threatens the validity of \llm{} evaluations. 
In this section, we aim to investigate whether open-source \llms{}, such as \codebert{}, \graphcodebert{}, and \unixcoder{}, are also vulnerable to such~issue.

A straightforward way to verify this would be to replicate the pre-training of these open-source models and apply a similar approach as before. However, this approach presents two key challenges: 1) the pre-training procedures for these models are not publicly available (as is the case for \codebert{}, \graphcodebert{}, and \unixcoder{}~\cite{feng2020codebert,guo2020graphcodebert,ahmad2021unified}), and 2) training these models with the full \csn{} and the original hyperparameters demands substantial GPU resources; for example, \codebert{} was pre-trained using 16 interconnected NVIDIA Tesla V100 GPUs.

Thus, we adopt an alternative approach that does not require pre-training the models from scratch. Specifically, we first identify a property $P$ that is indicative of models affected by data leakage. This property is present in leaky \llms{} and absent in those not affected by leakage.

The test set of a fine-tuning dataset can be divided into two groups: one that is included in the pre-trained dataset (\textit{leaky test group}) and one that is not included (\textit{non-leaky test group}). When the layers of the encoder model are frozen, 
it is likely to exhibit
decrease in performance for both groups, given that less parameters of the model are adapted to the fine-tuning dataset and task. 

However, for the leaky \llm{}, this performance decrease would be much faster on the non-leaky test group compared to the leaky test group. On the other hand, for the non-leaky \llm{}, the rate of performance decrease would be similar for both groups. This occurs because, during pre-training, the leaky \llm{} acquires some knowledge about the leaky test samples. When the layers are frozen, this prior knowledge is retained, which prevents rapid performance degradation in the leaky test group.

To illustrate this phenomenon, we consider our pre-trained encoder \llms{} and the code search task using the \pydataset{} dataset. We then run four linear regression models where $X$ represents the number of frozen layers and $Y$ represents the performance of a given test group. Table~\ref{tab:slope} displays the slopes for each \llm{} type and test group. It can be observed that, for the leaky \llm{}, the performance decreases more rapidly for the non-leaky group (slope of $-0.41$) compared to the leaky group (slope of $-0.25$) as more layers are frozen. In contrast, for the non-leaky \llm{}, the performance degradation is similar for both groups (slopes of $-0.37$ and $-0.38$ for the leaky and non-leaky groups, respectively).

\begin{table}[]
\caption{Regression slopes for each \llm{} type and test group.}
\label{tab:slope}
\centering
\begin{tabular}{lcc}
\hline
                & Leaky group & Non-leaky group \\ \hline
Leaky \llm{}     & -0.25         & -0.41             \\
Non-leaky \llm{} & -0.37         & -0.38             \\ \hline
\end{tabular}
\end{table}

\begin{figure*}[h]
\centering
\begin{subfigure}{.5\textwidth}
  \centering
  \includegraphics[width=.8\linewidth]{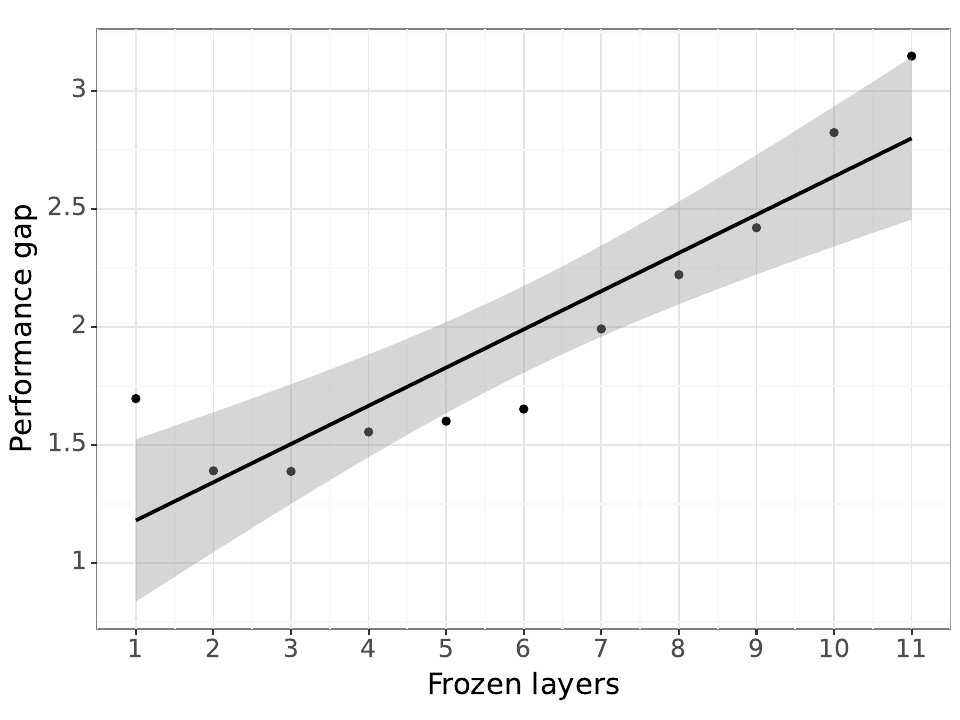}
  \caption{Correlation plot for the leaky \llm{}.}
  \label{fig:sub1}
\end{subfigure}%
\begin{subfigure}{.5\textwidth}
  \centering
  \includegraphics[width=.8\linewidth]{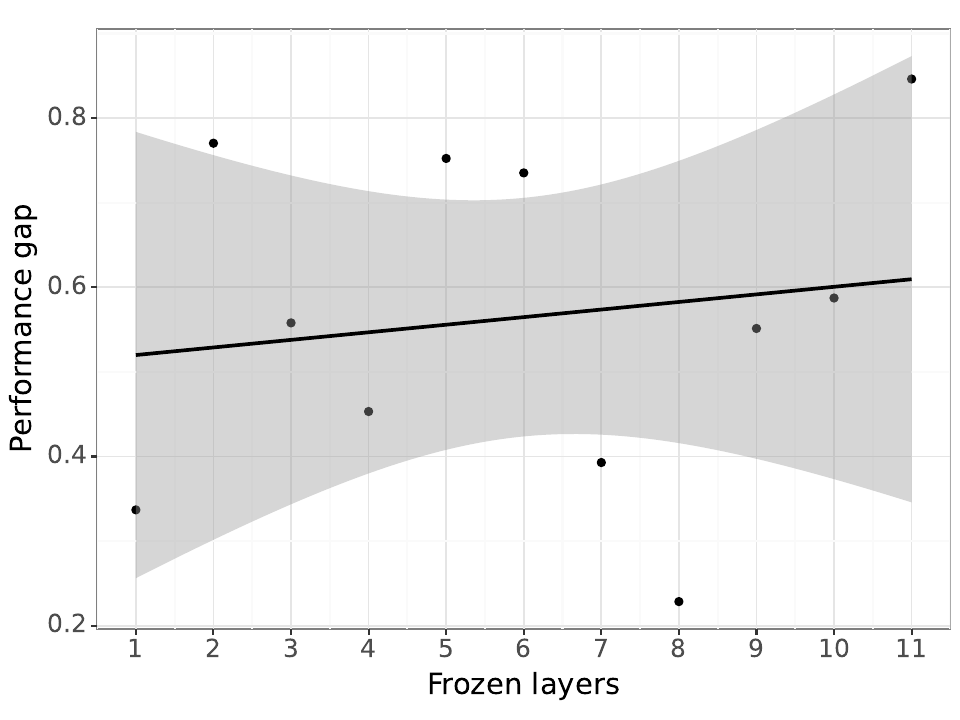}
  \caption{Correlation plot for the non-leaky \llm{}.}
  \label{fig:sub2}
\end{subfigure}
\caption{Correlation plot for each type of \llm{}.}
\label{fig:corrplots}
\end{figure*}

An equivalent way to visualize and quantify this effect is by measuring the correlation between the number of frozen layers and the performance gap of the given \llm{} between the leaky and non-leaky test portions. For the leaky \llm{}, a positive correlation would be observed since the performance of the non-leaky group decreases much faster than that of the leaky group, resulting in an increasing performance gap as more layers are frozen. Conversely, for the non-leaky \llm{}, there should be little to no correlation.

To illustrate this idea, Figs.~\ref{fig:sub1} and \ref{fig:sub2} show the correlation plots between the number of frozen layers and the performance gaps between the two test groups for each \llm{} type. For the leaky \llm{}, there is a strong correlation (Spearman coefficient of $0.85$, \pvalue{} $< 0.01$). In contrast, for the non-leaky \llm{}, there is no statistically significant correlation (Spearman coefficient of $0.16$, \pvalue{} $=0.63$).

We will now use this correlation method to demonstrate that \codebert{}, \graphcodebert{}, and \unixcoder{} could be affected by the inter-dataset code duplication threat. These models have been pre-trained with \csn{}. Therefore, the leaky and non-leaky test groups will be the same as in our previous experiments, allowing us to draw correlation plots and compute the correlation coefficients.

Fig.~\ref{fig:corrpretrained} shows the correlation plots for the three pre-trained models. The Spearman correlation coefficients are $0.91$, $0.80$, and $0.67$ for \unixcoder{}, \graphcodebert{}, and \codebert{}, respectively, all with associated \pvalue{}s $< 0.05$. The strongest correlation observed for \unixcoder{}, compared to the other two models, could be explained by its pre-training tasks. \unixcoder{} includes a pre-training task (the code fragment representation learning task~\cite{guo2022unixcoder}) that explicitly aligns code snippets with their documentation, which is not the case for \codebert{} and \graphcodebert{}. As for the difference between \codebert{} and \graphcodebert{}, it may be attributed to the fact that \graphcodebert{} was initialized with \codebert{} weights before starting the pre-training. In other words, \graphcodebert{} was trained for more epochs on \csn{}, potentially leading to greater memorization of the leaky test samples.

\begin{tcolorbox}
{\color{black}
Positive correlations are observed between the performance gap of the leaky group and the non-leaky one, and the number of frozen layers in the case of \codebert{}, \graphcodebert{}, and \unixcoder{}. Consequently, there is evidence that these open-source \llms{} could be affected by the inter-dataset code duplication threat.
}
\end{tcolorbox}

\begin{figure}[t]
    \centerline{\includegraphics[width=0.45\textwidth]{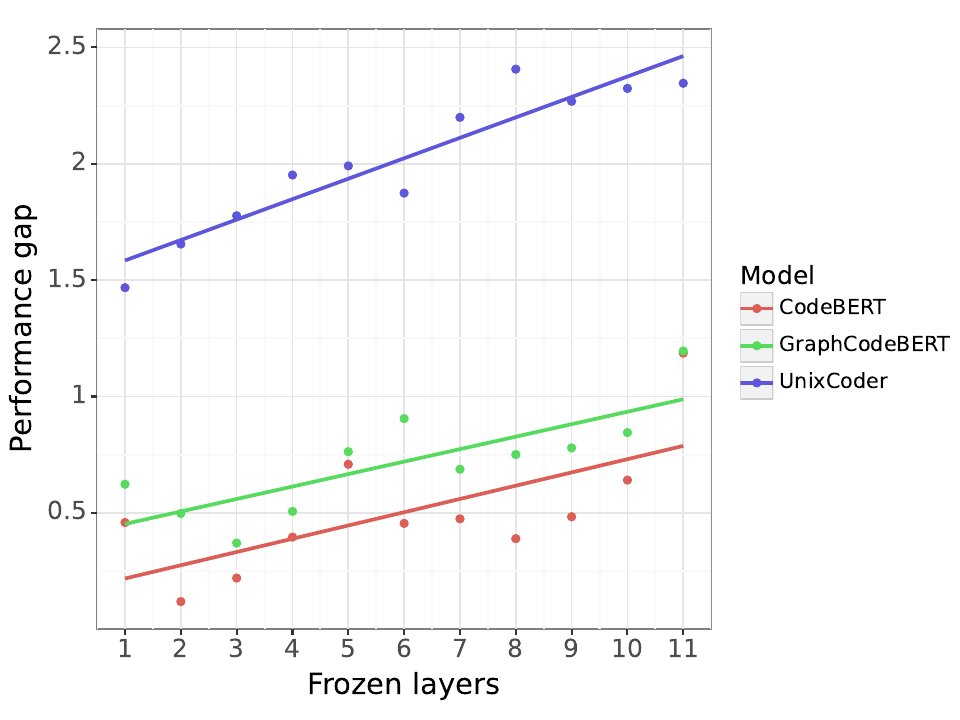}}
    \caption{Correlation plots for \codebert{}, \graphcodebert{}, and \unixcoder{}. The $x-$axis is the number of frozen layers and the $y-$axis is the performance gap between the leaky group and the non-leaky one.}
    \label{fig:corrpretrained}
\end{figure}

}

\section{Discussion}
\label{sect:discussion}

{\color{black}
Here we discuss key implications of the study and outline the connection 
between the effects of inter-dataset code duplication on neural networks with catastrophic forgetting.
We also present data points encountered during pre-training that were either remembered, partially remembered, or not remembered after fine-tuning. Finally, we discuss aspects that may pose risks to the experiment's validity.
}

\subsection{Key implications}

In this work, we demonstrate the presence of the inter-dataset code duplication phenomenon. Additionally, we find that this phenomenon poses a threat to the evaluation across various \llms{}, tasks, and datasets. Consequently, we summarize the main takeaways from this paper below.

\begin{tcolorbox}[colback=blue!20,coltext=black]
\textbf{Takeaway 1.} Always assess the overlap between the pre-training and the fine-tuning datasets.
\end{tcolorbox}

We have illustrated a non-trivial overlap in several fine-tuning datasets with a widely used pre-training dataset. It is logical to assume that larger pre-training datasets are more likely to have increased overlap with fine-tuning datasets. In this context, when releasing a new pre-training dataset, it could be beneficial to provide a tool for determining if a given piece of code is included in such a dataset. {\color{black} The \textsc{Stack} dataset~\cite{kocetkov2022stack}, an open-source pre-training dataset comprising over 6TB of permissively-licensed source code files extracted from GitHub, adheres to this practice. It includes a bloom filter-based tool to identify whether a given code snippet is part of this extensive dataset~\cite{marone2023dataportraits}.
}

\begin{tcolorbox}[colback=blue!20,coltext=black]
\textbf{Takeaway 2.} If overlapping is detected, remove the samples from the testing set used previously during pre-training to avoid data leakage and achieve unbiased performance evaluations.
\end{tcolorbox}

This practice is not widespread. For instance, Zou \etal{}~\cite{zhou2021assessing} assesses generalizability of \codebert{} using datasets distinct from the \csn{} dataset. However, one of the datasets employed, \pydataset{}, exhibits non-trivial overlap with \csn{}, as demonstrated in our experiments.

\begin{tcolorbox}[colback=blue!20,coltext=black]
\textbf{Takeaway 3.} Avoid fine-tuning a new \llm{} with a subset of the pre-training dataset to ensure a fair comparison with other baselines.
\end{tcolorbox}

{\color{black}
\codebert{}, \graphcodebert{}, \unixcoder{}, and \codetfive{}~\cite{feng2020codebert,guo2020graphcodebert,guo2022unixcoder,wang2021codet5}
do not adhere to the above takeaway. 
We show that \llms{} generally exhibit improved performance on samples encountered during pre-training (due to the fact that they have been seen during pre-training). Consequently, the reported figures in such studies may be inflated, and comparisons with other \llms{} used as baselines could be unfairly skewed. One way to tackle such issue is to deduplicate the initial (pre-training) dataset and split it into the pre-training dataset and the fine-tuning one.}

\begin{tcolorbox}[colback=blue!20,coltext=black]
\textbf{Takeway 4.} If removing overlapping is not possible, 
{\color{black} lightweight fine-tuning approaches should be used with caution, as they are more susceptible to the inter-dataset code duplication threat.}
\end{tcolorbox}

Our paper emphasizes that the impact of inter-dataset code duplication is potentially more pronounced when using lightweight fine-tuning techniques (LoRA, prefix tuning, and layer freezing). Related to this and in the context of software engineering, there are studies that explore lightweight fine-tuning approaches~\cite{shi2023towards,ayupov2022parameter}, and their reported findings may be influenced by the insights and findings of the current paper. {\color{black} For instance, Ayupov and Chirkova~\cite{ayupov2022parameter} investigate lightweight fine-tuning techniques, such as LoRA, and fine-tune \codetfive{} for code summarization, employing \csn{}. Shi \etal{}~\cite{shi2023towards} propose an efficient fine-tuning method for pre-trained code models through layer freezing. In their experiments, the authors fine-tune the frozen versions of \unixcoder{} and \graphcodebert{} for code search, once again using \csn{}.}

\subsection{Relation with catastrophic forgetting}

Neural networks are susceptible to the phenomenon of catastrophic forgetting, where training the network for a new task erases knowledge gained from previous tasks~\cite{mccloskey1989catastrophic}. In the context of pre-training and fine-tuning, a similar setup is employed. Initially, the model undergoes pre-training on a dataset with various tasks, followed by fine-tuning on a smaller dataset for a specific task.

{\color{black} Previous research has demonstrated that mitigating the effects of catastrophic forgetting can be achieved by freezing a significant subset of parameters~\cite{peng2018overcoming}. Our findings align with these observations. Specifically, we observe that, when employing lightweight fine-tuning approaches such as freezing the model, LoRA, or prefix tuning, the models are more prone to the inter-dataset code duplication phenomenon.}



{\color{black}
\subsection{Case studies}

\begin{table*}[h]
\caption{Examples of fully remembered (FR), partially remembered (PR), and not remembered (NR) instances.}
\label{tab:examples}
\centering
\begin{tabular}{llll}
\hline
                     & \textsc{Input code snippet} & \textsc{Generated summary} & \textsc{Ground truth} \\ \hline
FR     &     
\lstset{style=mystylejava}
\begin{lstlisting}
public int read() throws IOException {
  int b0=fInputStream.read();
  if (b0 > 0x80) {
    throw new IOException(Localizer.getMessage(
    "jsp.error.xml.invalidASCII",
    Integer.toString(b0)));
  }
  return b0;
}
\end{lstlisting}             &  

\parbox{4cm}{read a single character. this method will block until a character is available, an i$/$o error occurs, or the end of the stream is reached. $<$p$/$$>$$<$p$>$ subclasses that intend to support efficient single-character input should override this method. \\
}
&
\parbox{4cm}{read a single character. this method will block until a character is available, an i$/$o error occurs, or the end of the stream is reached. $<$p$>$ subclasses that intend to support efficient single-character input should override this method. \\
}

\\ \hline
PR &  
\lstset{style=mystylejava}
\begin{lstlisting}
private static Properties loadProperties(
String propertiesFile){
  Properties properties=new Properties();
  try (InputStream is=new FileInputStream(propertiesFile)){
    properties.load(is);
  }
 catch (  IOException e) {
    throw new RuntimeException("failed to load properties",e);
  }
  return properties;
}
\end{lstlisting}  

&   
\parbox{4cm}{loads properties from a properties file.
}

& 
\parbox{4cm}{loads properties from a properties file on the local filesystem.
}

\\ \hline
NR       &  

\lstset{style=mystylejava}
\begin{lstlisting}
private void paintForegroundEnabled(Graphics2D g,
int width,int height){
  Shape s=decodeArrowPath(width,height);
  g.setPaint(enabledColor);
  g.fill(s);
}
\end{lstlisting}     

&       

\parbox{4cm}{paint the background of the button using the specified colors.
}

&       
\parbox{4cm}{
paint the arrow in enabled state.
}

\\ \hline
\end{tabular}
\end{table*}

We now provide examples of code samples seen during pre-training categorized based on their familiarity to the fine-tuned \llm{}: fully remembered, partially remembered, and not remembered. These examples are drawn from our fine-tuned leaky \llm{} using \tlc{} and the corresponding leaky test portion of the dataset.

To categorize a given sample, we use the \bleu{} score~\cite{papineni2002bleu} of the generated summary compared to the ground truth, as it is highly dependent on token overlap.
We consider a sample to be \emph{fully remembered} by the \llm{} if its \bleu{} score exceeds 60, indicating \textit{quality often better than humans} according to Google's interpretation table~\cite{shi2022evaluation}. A sample is considered \emph{partially remembered} if its \bleu{} score falls between 30 and 60, which represents \textit{reasonably good translation}. If a sample's \bleu{} score is below 10, it is considered as \emph{not remembered} (\textit{useless translation}).

Table~\ref{tab:examples} presents an example from each category\footnote{More examples can be found in the GitHub repository of our replication package.}. The documentation generated from the fully remembered sample differs from the ground truth in only one small substring ($<$p$/$$>$). It is likely that the association between the input code snippet and its documentation was memorized during pre-training, as the latter part of the documentation (\ie{} \textit{subclasses that intend to [...]}) is very difficult to infer solely from the input code snippet without further context. In the partially remembered sample, we observe that the generated summary is nearly identical to the ground truth, but the leaky \llm{} omits the last part of the summary (\ie{} \textit{on the local filesystem}). Finally, 
there are instances where the leaky \llm{} (catastrophically) forgets the association learned during pre-training and produces incorrect documentation, as shown in the last row of the table.
}

\subsection{Threats to validity}

\subsubsection{Internal validity}
\llm{} fine-tuning can be non-deterministic due to the random initialization and shuffling. Consequently, the evaluation result for each fine-tuning setup may differ. To minimize this effect, we conduct several fine-tuning procedures for the same setup, each with different random seeds. 

\subsubsection{External validity}

{
\color{black}
We identify four primary external threats to validity.
The first threat is associated with the \emph{limited consideration of programming languages} in this paper, specifically Java and Python. While we hypothesize that these effects may be applicable to other programming languages, an exploration of this possibility is deferred to future research. The second threat is related to the \emph{representativeness of the considered code datasets and code tasks}. We mitigate the threat by using \csn{}~\cite{husain2019codesearchnet}, a widely-used pre-training dataset. Additionally, we incorporate various representative fine-tuning datasets and code tasks employed in prior \se{} research~\cite{wan2022naturalcc, lu2021codexglue, hu2018summarizing, leclair2019neural}.
The third threat relates to the \emph{pre-training tasks}. Although the literature describes numerous pre-training tasks~\cite{tufano2023automating}, we have experimented with only two, chosen based on popular \llms{} used in software engineering. We believe that the type of pre-training task may naturally influence the inter-dataset code duplication threat, and we leave further investigation of this for future work. Finally, the last threat concerns the \emph{model architectures}. This paper experiments only with encoder-only and encoder-decoder models. Therefore, the conclusions drawn may not apply to decoder-only models, and further investigation is needed.
}


\subsubsection{Construct validity}

Choosing the performance metrics for assessing performance in downstream tasks could pose a threat to construct validity. To address this concern, we opt for \mrr{} and \bleu{} metrics, which have been employed in prior studies to evaluate \llms{} in the tasks examined in this paper~\cite{lu2021codexglue,ahmed2022multilingual,feng2020codebert,guo2020graphcodebert,husain2019codesearchnet}.



\section{Conclusion \& Future work}
\label{sect:conclusion}

{\color{black} In this paper, we have demonstrated the existence of the \textit{inter-dataset code duplication} phenomenon and its potential negative impact on the reliability of \llms{} evaluations in SE tasks. Furthermore, we conclude that this impact is accentuated by the chosen fine-tuning technique. Specifically, lightweight fine-tuning techniques are more vulnerable to the identified~threat. Furthermore, we find evidence that popular open-source \llms{} (\codebert{}, \graphcodebert{}, and \unixcoder{}) are affected by this threat.}

In future work, we aim to extend this study to include additional programming languages and \se{} tasks. In this paper, our primary focus has been on encoder-only and encoder-decoder language models. Given the rising popularity of open source decoding-only language models (\eg{} \textsc{StarCoder}~\cite{li2023starcoder}, \textsc{SantaCoder}~\cite{allal2023santacoder}, \textsc{GPT-Neox}~\cite{black2022gpt}, etc.), we also plan to investigate these effects in such \llms{}. {\color{black} 
However, due to the immense size of the current decoder-only pre-training datasets, which are several orders of magnitude larger than \csn{}, computing intersections between the fine-tuning and pre-training datasets is challenging. Therefore, to study the inter-duplication phenomenon in these cases, we plan to use alternative technologies such as data portraits~\cite{marone2023dataportraits} instead of the Allamanis tool~\cite{allamanis2019adverse}.}



\ifCLASSOPTIONcompsoc
  \section*{Acknowledgments}
\else
  \section*{Acknowledgment}
\fi

This work was partially supported by the Wallenberg AI, Autonomous Systems and Software Program (WASP) funded by the Knut and Alice Wallenberg Foundation; and by the grant TED2021-129381B-C22 (SATORI project) funded by MCIN/AEI/10.13039/501100011033.

\ifCLASSOPTIONcaptionsoff
  \newpage
\fi

\bibliographystyle{IEEEtran}
\bibliography{database.bib}

\end{document}